\documentclass[journal]{IEEEtran}
\usepackage{amsmath,amssymb,amsfonts}
\usepackage{booktabs}
\usepackage{spverbatim}
\usepackage{graphicx}
\usepackage{hyperref}
\usepackage{xurl}
\usepackage{tabularx}

\usepackage{refcount}
\newcounter{fncntr}
\newcommand{\fnmark}[1]{\refstepcounter{fncntr}\label{#1}\footnotemark[\getrefnumber{#1}]}
\newcommand{\fntext}[2]{\footnotetext[\getrefnumber{#1}]{#2}}

\begin {document}
\title{Why network segmentation projects fail}

\author{\IEEEauthorblockN{Rohit Dube}
\IEEEauthorblockA{\textit{Cisco Systems Inc.} \\
                 170 West Tasman Drive, San Jose, CA 95134 USA \\
                 }
}

\maketitle

\begin{abstract}
Network segmentation is a foundational enterprise security control.
Despite its recognized benefits, segmentation initiatives frequently fail in practice, and the field lacks a systematic empirical explanation for why these projects do not achieve their intended outcomes.
This paper presents an empirical study of failed segmentation projects based on a survey of 400 U.S.-based\ network security practitioners.
The survey was grounded in a two-part failure framework that separately measures general IT project failure factors and segmentation-specific technical and operational barriers.
Clustering analysis of the responses reveals four distinct failure archetypes, ranging from projects in which every factor contributed simultaneously to projects where governance was comparatively better but specific segmentation challenges proved decisive.
The archetypes correspond to real differences in how segmentation was attempted: projects that included campus networks were more likely to experience the broadest or most technically intense failures.
On the other hand, the archetypes did not differ significantly by workload type.
\end{abstract}


\section {Introduction} \label{sec:intro}

Network segmentation is widely regarded as a foundational control in modern enterprise cybersecurity architectures because it reduces attack surface, constrains lateral movement, and enables more granular enforcement of security policy.
Prior research demonstrates that segmented networks significantly outperform flat networks in resisting malware propagation and unauthorized access, with empirical studies showing that simulated attacks are contained entirely within isolated segments rather than spreading across enterprise infrastructure \cite{bredesen2025network}.
Similarly, academic and practitioner-oriented analyses emphasize segmentation as a core defense-in-depth mechanism that supports least-privilege access, breach containment, and regulatory compliance, even as it introduces additional operational and management complexity \cite{kotha2020network}.
At the same time, prior work highlights that while segmentation is clearly beneficial, guidance on designing and applying effective segmentation architectures remains vague, forcing practitioners to rely heavily on judgment when making segmentation decisions \cite{wagner2016towards}.

Practitioners consistently report that network segmentation initiatives often fail to achieve their intended outcomes or are abandoned before completion.
These failures are well known anecdotally in industry practice~\cite{columbus2022forrester, elisity2026}, yet neither the academic literature nor practitioner-focused publications provide a systematic, empirical examination of why segmentation deployment projects fail.

This paper addresses that gap by presenting an empirical study based on a survey of network security practitioners with direct experience in failed segmentation projects.
Practitioner perceptions are both a valid and important source of evidence, as these practitioners are the actors making design and implementation decisions during segmentation projects.
A survey-based approach was chosen to gather practitioner views over alternatives such as interviews or focus groups because a survey enables a sample size large enough to support statistical analysis of failure patterns.
The study captures perceptions across project management and technical dimensions of failure and applies quantitative clustering techniques to identify failure archetypes.

The analysis reveals four distinct failure archetypes---\textit{Perfect Storm}, \textit{Diffuse Friction}, \textit{Operational Drag}, and \textit{Scope \& Visibility Trap}---ranging from projects where every factor contributed simultaneously to projects where governance was comparatively better but specific segmentation challenges proved decisive.
The archetypes are associated with distinct project characteristics: projects that included campus networks were more likely to experience the broadest (Perfect Storm) or most technically intense (Scope \& Visibility Trap) failures, and the most technically intense archetype was further associated with traditional Layer-2 macro-segmentation, while the archetypes did not significantly differ by workload type or failure type.
Because the archetypes correspond to how segmentation was attempted rather than to what was being segmented, they carry direct implications for practice: a project's risk profile can be anticipated from its scope and approach before failure occurs, and recovery cannot follow a single template, since a pervasive failure and a narrowly technical one call for different interventions.

The remainder of this paper is organized as follows.
Section~\ref{sec:related} reviews prior academic and practitioner work relevant to IT project failure and network segmentation challenges.
Section~\ref{sec:it_failure_framework} introduces a general framework for IT project failure and describes how its constructs are measured in an empirical study.
Section~\ref{sec:seg_failure_framework} extends this framework to include the unique considerations of segmentation projects and describes the corresponding measurement constructs.
Section~\ref{sec:archetypes} develops network segmentation failure archetypes from the survey results.
Section~\ref{sec:what} analyzes practitioner-proposed solutions to segmentation failures.
Section~\ref{sec:limitations} points out the limitations of our work, and Section~\ref{sec:conclusion} concludes with a summary of our work.

\section {Related Work} \label{sec:related}

A substantial body of research has examined why IT projects fail across different organizational contexts.
Early empirical work demonstrated that project failure is associated with a set of managerially controllable factors whose relative importance varies by project type and lifecycle stage, challenging the view that failures are idiosyncratic or purely technical \cite{pinto1990causes}.
Large-scale empirical studies further show that cost overruns, delays, and benefit shortfalls are not explained by project complexity alone but instead reflect planning and governance deficiencies \cite{budzier2013overspend}.
Research focusing on practitioner perceptions highlights how failures emerge through causal chains, with communication and cooperation failures acting as bridge causes across dimensions \cite{lehtinen2014perceived}.
In-depth investigations of failed government IT projects find that most root causes originate outside of programming activities and instead reflect deficiencies in early problem formulation, requirements understanding, and organizational decision-making \cite{lauesen2020project}.
Synthesis work spanning more than three decades confirms strong consistency in these findings over time \cite{schmidt2023mitigating}, while studies of project complexity emphasize that failure factors cluster, reinforcing the need for multidimensional frameworks \cite{montequin2018exploring}.

Industry reports provide consistent evidence that segmentation initiatives encounter failure mechanisms distinct from general IT project challenges and rooted in the technical and operational properties of modern enterprise environments.
A large-scale industry survey identifies a convergent set of technical barriers---architectural complexity, insufficient asset visibility, difficulty mapping communication flows, manual policy burden, and tooling limitations---as the primary segmentation challenges \cite{Cisco2025SegmentationReport}.
The same study shows that segmentation efforts frequently stall despite high organizational prioritization, with only one-third of respondents reporting full implementation of both macro- and micro-segmentation, indicating that failure may arise from incomplete environmental understanding rather than insufficient intent or governance \cite{Cisco2025SegmentationReport}.
Independent industry analysis reinforces these conclusions, reporting that segmentation initiatives commonly stall at the pilot or partial deployment stage due to concerns about policy safety and application outages, even though segmentation is widely recognized as strategically important \cite{Akamai2025SegmentationImpactStudy}.
Earlier industry survey results similarly note that lack of internal expertise and perceived operational risk slow segmentation progress, leading many organizations to limit enforcement scope despite acknowledging segmentation's effectiveness \cite{Akamai2023StateOfSegmentation}.
Taken together, these industry studies motivate the need for a dedicated segmentation project failure framework distinct from general IT project failure models.

Prior work has established the security benefits of network segmentation through analytical studies, simulations, and experimental evaluations, while also noting the practical difficulty of selecting, designing, and maintaining segmentation architectures \cite{bredesen2025network,kotha2020network,wagner2016towards}.
Recent taxonomy work highlights the lack of empirical studies examining segmentation outcomes in real deployment and organizational settings \cite{dube2026taxonomy}.
Complementary formal-methods research further highlights that segmentation guidance is often vague or underspecified, making it difficult to apply consistently or scale to large, complex environments without relying heavily on expert judgment \cite{mhaskar2021formal}.
This study addresses the gaps by translating the identified challenges into concrete failure factors, examining segmentation project failures from practitioner perspectives, and connecting segmentation outcomes to both general IT project failure causes and segmentation-specific challenges.

\section {General IT Project Failure Framework} \label{sec:it_failure_framework}

In this section, we synthesize prior findings on general IT project failures into a concise framework suitable for empirical measurement.
The framework emphasizes general, managerially controllable factors that recur across project types and remain stable over time despite changes in technology.

The first dimension of the framework concerns strategic alignment and governance.
Failures in this dimension originate from unclear project goals and insufficient leadership sponsorship or decision-making authority.
Weak alignment at project initiation creates ambiguity around priorities and success criteria, which can constrain effective planning and coordination.
Governance deficiencies further limit the organization's ability to allocate resources, resolve conflicts, and respond to emerging risks during delivery \cite{pinto1990causes,schmidt2023mitigating}.

The second dimension addresses requirements and scope management.
Unstable requirements disrupt estimation, sequencing, and coordination, and often propagate execution problems even when governance structures are nominally in place.
Prior research shows that requirements-related issues frequently act as connective mechanisms linking strategic weaknesses to execution breakdowns \cite{lehtinen2014perceived,lauesen2020project}.

The third dimension combines project planning, execution, and control.
This dimension captures both the realism of initial feasibility assumptions and the effectiveness of ongoing managerial oversight during delivery.
Unrealistic schedules or resource assumptions increase exposure to failure by establishing infeasible baselines \cite{budzier2013overspend}.
However, failure often materializes through inadequate monitoring, delayed risk response, or ineffective corrective action rather than through planning deficiencies alone \cite{schmidt2023mitigating}.

The final dimension concerns coordination and communication among stakeholders.
Unlike the preceding dimensions, coordination and communication tend to permeate all phases of the project lifecycle.
Communication breakdowns amplify the effects of governance gaps, unstable requirements, and execution challenges by impeding shared understanding and timely decision-making.
As project complexity and interdependence increase, coordination failures become more frequent and more consequential \cite{lehtinen2014perceived,montequin2018exploring}.

Taken together, these dimensions form a concise and conceptually coherent framework for general IT project failure.
The framework reflects a progression from strategic foundations, through definition and delivery, while explicitly accounting for universal coordination mechanisms.
Table~\ref{tab:failure_framework_mapping} summarizes the mapping between the framework dimensions and prior literature used to ground this synthesis.
This framework provides a theoretical basis for the quantification and empirical analysis presented in subsequent sections.

To make these dimensions empirically tractable, we operationalize each as one or two Likert-scale\fnmark{likert} survey items as shown in Table~\ref{tab:survey_items} \cite{likert1932technique}.
Items B1 and B2 measure strategic alignment and governance, capturing goal clarity and leadership sponsorship respectively.
Item B3 measures requirements and scope management through experienced scope creep and changing requirements.
Items B4 and B5 jointly measure project planning, execution, and control: B4 captures the realism of timelines and resources, while B5 captures the effectiveness of ongoing risk identification and corrective action during delivery.
Item B6 measures coordination and communication among stakeholders.

\fntext{likert}{
  All items used a five-point agreement scale: 1 = Strongly disagree, 2 = Slightly disagree, 3 = Neither agree nor disagree, 4 = Slightly agree, 5 = Strongly agree.
}

\begin{table*}[htbp]
\caption{Mapping of IT Project Failure Framework Dimensions to Prior Literature.}
\label{tab:failure_framework_mapping}
\centering
\begin{tabular}{p{4.2cm} p{6.2cm} p{6.2cm}}
\hline
\textbf{Framework Dimension} & \textbf{Key Failure Aspects} & \textbf{Representative Studies} \\
\hline

Strategic Alignment \& Governance &
Unclear goals, weak sponsorship, insufficient decision authority &
Pinto and Mantel~\cite{pinto1990causes};
Schmidt~\cite{schmidt2023mitigating} \\

\hline

Requirements \& Scope Management &
Scope creep, changing or unclear requirements &
Lehtinen et al.~\cite{lehtinen2014perceived};
Lauesen~\cite{lauesen2020project};
Schmidt~\cite{schmidt2023mitigating} \\

\hline

Project Planning, Execution \& Control &
Unrealistic schedules, resource misalignment, inadequate monitoring, weak risk management, delayed corrective action &
Pinto and Mantel~\cite{pinto1990causes};
Budzier and Flyvbjerg~\cite{budzier2013overspend};
Schmidt~\cite{schmidt2023mitigating};
Montequ\'{\i}n et al.~\cite{montequin2018exploring} \\

\hline

Coordination \& Communication &
Poor stakeholder coordination, ineffective communication &
Lehtinen et al.~\cite{lehtinen2014perceived};
Montequ\'{\i}n et al.~\cite{montequin2018exploring} \\

\hline
\end{tabular}
\end{table*}

\begin{table}[htbp]
\caption{Survey Items Measuring General IT Project Failure Factors.}
\label{tab:survey_items}
\centering
\begin{tabular}{p{0.7cm} p{4.1cm} p{2.3cm}}
\hline
\textbf{Item} & \textbf{Survey Question (Agreement Scale)} & \textbf{Framework Dimension} \\
\hline

B1 & Goals were unclear or inconsistently defined at the outset. & Strategic Alignment \& Governance \\
\hline

B2 & Senior leadership sponsorship and project leadership were insufficient. & Strategic Alignment \& Governance \\
\hline

B3 & The project experienced scope creep or changing requirements. & Requirements \& Scope Management \\
\hline

B4 & The project timeline was unrealistic given the available resources. & Project Planning, Execution, \& Control \\
\hline

B5 & Issues and risks were not identified, tracked, or addressed effectively during project execution. & Project Planning, Execution, \& Control \\
\hline

B6 & Coordination and communication among stakeholders were ineffective. & Coordination \& Communication \\
\hline

\end{tabular}
\end{table}

\section{Segmentation Project Failure Framework} \label{sec:seg_failure_framework}

Building on the general IT project failure foundation, this section introduces a complementary framework that captures failure mechanisms specific to segmentation projects.
As summarized in the related work section (Section~\ref{sec:related}), industry studies consistently show that segmentation initiatives encounter technical and operational barriers that are not adequately explained by traditional IT project failure constructs.
Accordingly, the framework focuses on failure mechanisms arising from characteristics unique to segmentation.

The first framework component is \emph{architectural and environmental complexity}.
Segmentation projects are often implemented across on-premises data centers, cloud platforms, containers, and legacy systems.
Heterogeneity in infrastructure types complicates consistent policy design, deployment, and enforcement, increasing the likelihood of stalled or partial implementations \cite{Cisco2025SegmentationReport}.

The second component is \emph{insufficient visibility into assets and workloads}.
Segmentation depends on an accurate understanding of assets, workloads, and their roles within the environment.
When asset visibility is incomplete, segmentation policies become either overly permissive or operationally risky, undermining project objectives \cite{Cisco2025SegmentationReport}.

Closely related is the \emph{difficulty of identifying legitimate communication flows}.
Segmentation requires distinguishing required system-to-system communication from unnecessary or risky traffic.
Uncertainty in communication patterns leads to repeated policy revisions and prolonged pilot phases, limiting progress toward enforcement goals \cite{Cisco2025SegmentationReport}.

The framework further includes \emph{policy lifecycle and operational burden}.
Segmentation policies must evolve continuously as applications change and environments scale.
High manual effort in policy creation and maintenance increases operational burden and reduces long-term sustainability \cite{Cisco2025SegmentationReport}.

Another component is \emph{tooling maturity and automation limitations}.
Effective segmentation requires tooling that supports discovery, policy management, enforcement, and validation at scale.
Limitations in tooling maturity or integration increase operational overhead and constrain deployment scope \cite{Cisco2025SegmentationReport}.

Finally, the framework includes \emph{risk of business disruption}.
Segmentation misconfigurations can directly disrupt application availability or performance.
Perceived outage risk often constrains enforcement decisions, leading organizations to limit segmentation to lower-risk areas \cite{Akamai2023StateOfSegmentation}.

Each segmentation framework component is operationalized using a Likert-scale survey item as shown in Table~\ref{tab:segmentation_survey_items} \cite{likert1932technique}.
Item C1 captures architectural and environmental complexity, while C2 measures the degree to which insufficient asset visibility constrained policy design.
Item C3 assesses the difficulty of identifying legitimate communication flows between systems.
Item C4 evaluates whether excessive manual effort in policy creation and maintenance undermined sustainability, and C5 captures the extent to which tooling limitations contributed to failure.
Item C6 measures the degree to which concerns about application outages or business disruption constrained deployment.

\begin{table}[htbp]
\caption{Survey Items Measuring Segmentation Project Failure Factors.}
\label{tab:segmentation_survey_items}
\centering
\begin{tabular}{p{0.7cm} p{4.1cm} p{2.3cm}}
\hline
\textbf{Item} & \textbf{Survey Question (Agreement Scale)} & \textbf{Framework Dimension} \\
\hline

C1 & The complexity of the environment (e.g., hybrid cloud, containers, legacy systems) made segmentation difficult to implement successfully. & Architectural and Environmental Complexity \\
\hline

C2 & We lacked sufficient visibility into assets to design effective segmentation policies. & Insufficient Asset Visibility \\
\hline

C3 & Identifying legitimate communication flows between systems was more difficult than anticipated. & Difficulty Identifying Legitimate Communication Flows \\
\hline

C4 & Segmentation policies required excessive manual effort to create and maintain. & Policy Lifecycle and Operational Drag \\
\hline

C5 & Limitations or immaturity of segmentation tools (including third-party products) contributed significantly to project failure. & Tooling Maturity and Automation Limitations \\
\hline

C6 & Concerns about application outages or business disruption constrained how fully segmentation could be deployed. & Risk of Business Disruption \\
\hline

\end{tabular}
\end{table}

\section{Segmentation Failure Archetypes} \label{sec:archetypes}

\begin{figure}
  \centering
  \includegraphics[width=0.50\textwidth]{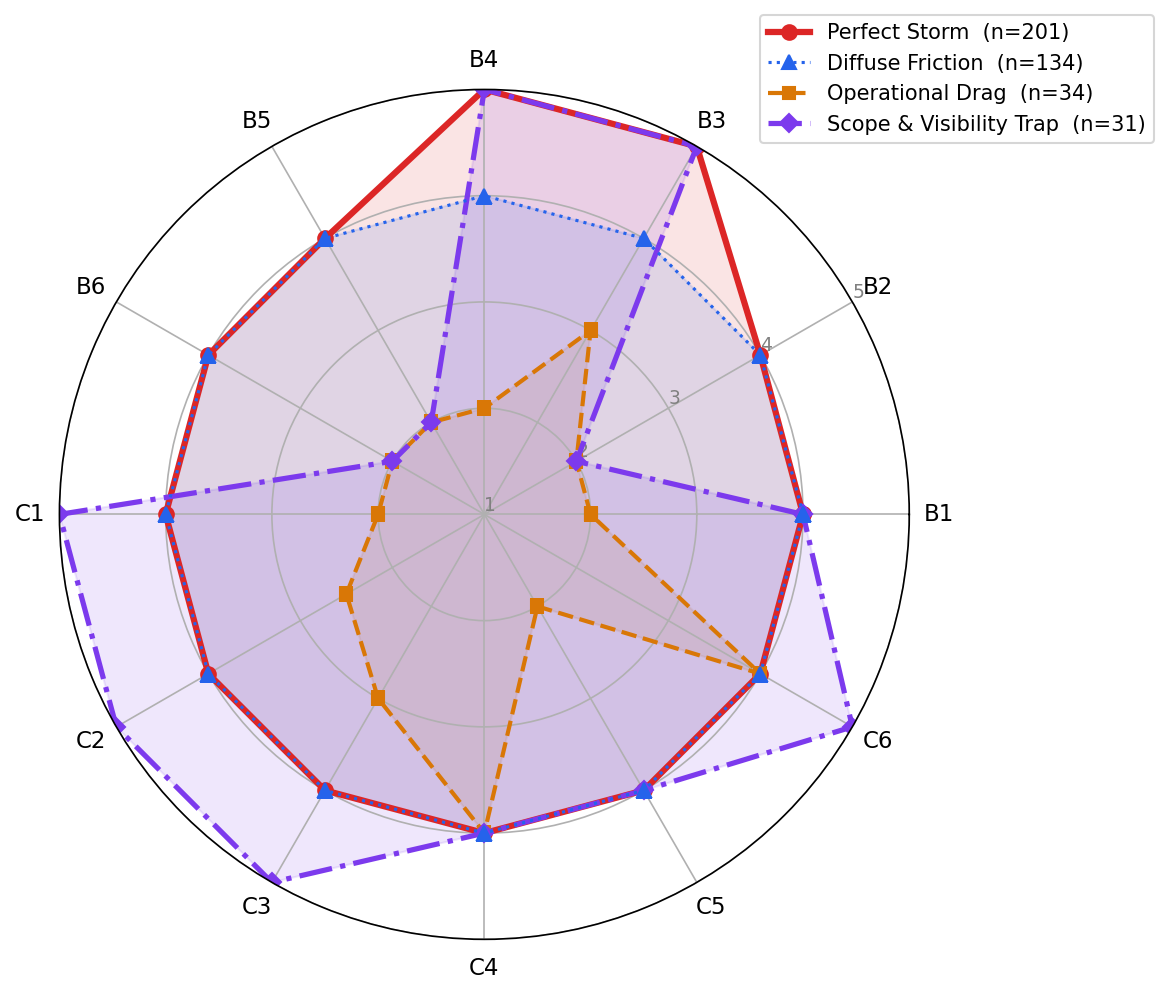}
  \caption{Radar plot of medians for LCA clusters.}
  \label{fig:lca_cluster}
\end{figure}

Figure \ref{fig:lca_cluster} previews the four failure archetypes identified by the analysis---\textit{Perfect Storm}, \textit{Diffuse Friction}, \textit{Operational Drag}, and \textit{Scope \& Visibility Trap}---each characterized by a different pattern of failure attribution across general IT project management factors (B factors) and segmentation-specific technical factors (C factors).
The remainder of this section describes how these archetypes were derived and what distinguishes them.

\subsection{Determining the Number of Clusters} \label{sec:determine_k}

\begin{table}[htbp]
\centering
\caption{Descriptive statistics for Likert-scale variables.}
\label{tab:likert_descriptives}
\begin{tabular}{lccccc}
\toprule
 & Min & Max & Median & Mean & SD \\
\midrule
B1 & 1 & 5 & 4.0 & 3.8 & 1.1 \\
B2 & 1 & 5 & 4.0 & 3.8 & 1.1 \\
B3 & 1 & 5 & 4.0 & 4.1 & 0.9 \\
B4 & 1 & 5 & 4.0 & 4.1 & 1.0 \\
B5 & 1 & 5 & 4.0 & 3.8 & 1.1 \\
B6 & 1 & 5 & 4.0 & 3.9 & 1.0 \\
C1 & 1 & 5 & 4.0 & 4.1 & 0.9 \\
C2 & 1 & 5 & 4.0 & 4.1 & 0.9 \\
C3 & 1 & 5 & 4.0 & 4.1 & 0.9 \\
C4 & 1 & 5 & 4.0 & 4.1 & 0.8 \\
C5 & 1 & 5 & 4.0 & 4.1 & 0.9 \\
C6 & 1 & 5 & 4.0 & 4.2 & 0.9 \\
\bottomrule
\end{tabular}
\end{table}

A survey containing the Likert-scale items and some related questions was conducted in late February and early March of 2026 and 400 responses from U.S.-based\ network security practitioners were obtained (see Appendix~\ref{sec:survey_admin},~\ref{sec:generalizability} for details).
Table~\ref{tab:likert_descriptives} summarizes the 12 items.
All items have a median of 4.0 and similar means (3.8--4.2), with C factors (C1--C6) marginally higher than B factors (B1--B6), yet the full 1--5 range is used for every item.
This combination of homogeneous aggregate statistics and substantial individual-level variability suggests that distinct subgroups with different failure attribution patterns may be present, motivating the use of a clustering technique (latent class analysis) that can recover them.

Latent Class Analysis (LCA) is a model-based clustering technique.
It assumes that observed survey responses come from a mixture of distinct, unobserved subgroups, each with its own response pattern.
For a given number of classes $K$, the model estimates two things: (i)~the probability of selecting each response option (1--5) on each Likert item for each class, and (ii)~the proportion of respondents in each class.
These parameters are estimated using the Expectation-Maximization (EM) algorithm~\cite{dempster1977maximum}.
The EM algorithm alternates between two steps:
  computing each respondent's probability of belonging to each class based on their responses (E-step),
  and updating each class's response probabilities to best explain the data (M-step).
Each respondent is assigned to the class with the highest posterior probability.
To guard against convergence to local optima, for each value of $K$ the EM algorithm was run 100 times with different random initializations, and the solution with the highest log-likelihood was retained for that model.\fnmark{stepmix}

\fntext{stepmix}{
  The fully unconstrained $K=4$ model estimates approximately 195 free parameters (48 response probabilities per class plus 3 mixing weights) for 400 observations.
  While this 2:1 observation-to-parameter ratio is low by structural equation modeling conventions, EM estimation with categorical indicators is more stable than the raw ratio suggests:
  the ordinal response structure implicitly constrains the parameter space, and the high entropy (0.904) and acceptable bootstrap stability of the solution (Appendix~\ref{sec:stability}) provide empirical confirmation that the model is not overfitting.
}

Unlike distance-based methods, LCA treats each Likert item as categorical and models the full distribution of responses within each class.
This makes it well suited for ordinal survey data~\cite{collins2010latent, vermunt2002latent}.\fnmark{lca} \fnmark{kmeans}

\fntext{lca}{
  LCA assumes conditional independence: within each class, the 12 indicators are assumed to be unrelated once class membership is accounted for.
  Some item pairs in the survey measure related constructs (e.g., B1 and B2 both capture governance; C1 and C2 both relate to environmental understanding), which could violate this assumption.
  Violations tend to produce additional classes that capture residual within-class correlations rather than genuine subgroups.
  The external validation results in Section~\ref{sec:external_validation}, which show significant associations between cluster membership and project characteristics not used in clustering, argue against a purely spurious solution.
}

\fntext{kmeans}{
  A more familiar alternative, $K$-means clustering, treats survey responses as continuous numbers and measures similarity using Euclidean distance.
  This implicitly assumes that the gap between ``Strongly Disagree'' (1) and ``Disagree'' (2) is the same as between ``Agree'' (4) and ``Strongly Agree'' (5), and that responses follow a bell-shaped distribution within each cluster, neither of which holds for five-point Likert scales.
  LCA avoids these assumptions by modeling the probability of each response option separately, making it the standard clustering method for ordinal survey data in the social and behavioral sciences.
}

Choosing the right number of classes requires fitting models for a range of $K$ values and comparing them.
Two commonly used measures are the Bayesian Information Criterion (BIC) and the Akaike Information Criterion (AIC).
Both weigh model fit against complexity, but BIC penalizes additional parameters more heavily and is often recommended as the primary guide for class enumeration~\cite{nylund2007deciding, nylund2018ten}.
Because BIC applies the strongest parsimony correction among common information criteria, it can favor simpler solutions even when additional classes capture meaningful structure.
In such cases, the rate of change in AIC provides a useful complement: when the improvement in AIC shrinks sharply, adding more classes offers diminishing returns~\cite{nylund2018ten,burnham2002model}.
Relative entropy measures how cleanly respondents are sorted into their assigned class, on a scale from 0 to 1, with values above 0.80 indicating high classification certainty and values below 0.60 suggesting poor separation~\cite{celeux1996entropy, ramaswamy1993empirical}.
Entropy should not be used as a selection criterion on its own, as it tends to increase mechanically with the number of classes~\cite{masyn2013latent}.
Finally, the size of the smallest class serves as a practical constraint: classes containing fewer than 5--8\% of the sample may lack sufficient observations to estimate class-specific parameters reliably and risk overfitting~\cite{nylund2018ten}.
Together, the AIC gradient, entropy, and minimum class size provide complementary evidence for class enumeration, particularly when BIC's strong parsimony penalty favors a solution simpler than domain knowledge would suggest.

Table~\ref{tab:lca_fit} shows fit indices for models with $K = 2$ through $K = 10$.
Based on the convergence of multiple criteria, $K = 4$ is selected as the optimal number of latent classes.
BIC rises monotonically from $K = 2$, driven by the large number of parameters each new class adds combined with the BIC penalty, which scales with $\log(N)$.
By itself, BIC would select $K = 2$, but the AIC trajectory tells a different story.
The largest AIC improvement occurs from $K = 2$ to $K = 3$, followed by a substantial further drop from $K = 3$ to $K = 4$.
From $K = 4$ to $K = 5$, the improvement shrinks considerably, and AIC begins rising at $K = 6$.
The region of diminishing returns spans $K = 3$ through $K = 5$, with $K = 4$ representing the point beyond which meaningful improvement largely ceases.
Entropy at $K = 4$ is 0.904, the highest of any solution in the $K = 2$ through $K = 5$ range and well above the 0.80 threshold for good classification certainty.
The $K = 5$ solution produces a smallest class of just 18 respondents (4.5\% of the sample), which falls below the 5--8\% range for adequate class size.
The $K = 4$ solution, by contrast, has a smallest class of 31 respondents (7.8\%), comfortably within this range.
While entropy continues to rise at higher values of $K$, those solutions lack support from the AIC curve and carry growing risk of unstable small classes.
Appendix~\ref{sec:stability} provides additional evidence, including a bootstrap stability analysis and likelihood ratio tests that jointly bracket the solution at $K = 4$.

\begin{table}[htbp]
\centering
\caption{Latent Class Analysis model fit indices for
$K = 2$ to $K = 10$ classes.}
\label{tab:lca_fit}
\begin{tabular}{cccccc}
\toprule
$K$ & AIC & $\Delta$AIC & BIC & Entropy &
Smallest Class (\%) \\
\midrule
2  & 11140.7 &            & 11623.6 & 0.877 & 31.0 \\
3  & 10924.6 & $-$216.1   & 11651.0 & 0.881 & 14.2 \\
4  & 10826.2 & $-$98.4    & 11796.1 & 0.904 & 7.8  \\
5  & 10806.2 & $-$20.0    & 12019.6 & 0.891 & 4.5  \\
6  & 10827.3 & $+$21.1    & 12284.2 & 0.908 & 4.5  \\
7  & 10853.2 & $+$26.0    & 12553.6 & 0.920 & 4.0  \\
8  & 10901.3 & $+$48.0    & 12845.1 & 0.899 & 3.8  \\
9  & 10942.3 & $+$41.1    & 13129.7 & 0.918 & 4.2  \\
10 & 11001.6 & $+$59.3    & 13432.4 & 0.903 & 3.0  \\
\bottomrule
\end{tabular}
\end{table}

\subsection{Cluster Signatures} \label{sec:cluster_signatures}

Tables~\ref{tab:cluster_medians} and~\ref{tab:cluster_endorsement} present, respectively, the median responses and endorsement rates (percentage rating an item $\geq 4$) for each of the four clusters.
The clusters are presented in order of decreasing size (number of respondents).
Figure~\ref{fig:lca_cluster} provides a complementary visual summary.

The largest cluster ($n = 201$, 50.2\%) exhibits the broadest failure attribution of any group.
Median responses are 4 or 5 on every item, and endorsement rates exceed 93\% for all 12 factors (Table~\ref{tab:cluster_endorsement}).
Both B factors and C factors are endorsed at comparable rates, with no meaningful variation across items.
In effect, the typical respondent in this cluster agrees or strongly agrees that every factor contributed to project failure.
We label this cluster \textit{Perfect Storm}: projects in which general IT project management failures and segmentation-specific technical challenges occurred simultaneously and pervasively.\fnmark{acquiescence}
A response-style check confirms that this broad endorsement reflects a substantive failure pattern rather than acquiescence bias (Appendix~\ref{sec:response_style}).

\fntext{acquiescence}{
  The broad endorsement in this cluster raises the question of acquiescence bias: a tendency to agree regardless of content.
  A response-style analysis (Appendix~\ref{sec:response_style}) shows that overall agreement level does not separate the archetypes, that within-respondent differentiation is lowest in Perfect Storm and highest in the more selective clusters, and that Perfect Storm membership is associated with external project characteristics not used in clustering, all of which argue against a response-style artifact.
}

The second-largest cluster ($n = 134$, 33.5\%) shows a similarly broad but less intense pattern.
Median responses are uniformly 4 across all 12 items, identical to much of the Perfect Storm cluster in Table~\ref{tab:cluster_medians}.
However, the endorsement rates reveal a clear difference.
B factors are endorsed at roughly 30 percentage points below the corresponding rates in Perfect Storm.
C factors are endorsed at higher rates than the B factors but still well below Perfect Storm.
Further, nearly two-thirds of respondents in this cluster rate C factors higher than B factors on average.
This cluster does not exhibit any single dominant failure cause.
Instead, the pattern suggests projects that accumulated moderate friction across multiple dimensions without a decisive point of collapse.
We label this cluster \textit{Diffuse Friction}.

The remaining two clusters break the pattern of broad endorsement.
Unlike Perfect Storm and Diffuse Friction, which endorse most or all failure factors, these clusters reject some or all B factors while endorsing specific C factors.

In the larger of the two ($n = 34$, 8.5\%), median responses on B factors are predominantly 2, with low endorsement rates for insufficient leadership (B2, 8.8\%) and unclear goals (B1, 17.6\%).
These respondents disagree that general project management problems caused their project to fail.
Two C factors stand out: excessive manual effort in policy creation and maintenance (C4, 55.9\%) and concerns about outages or business disruption (C6, 55.9\%), both with a median of 4.
This cluster describes projects where the operational burden of creating and maintaining segmentation policies was the primary barrier, compounded by reluctance to deploy aggressively due to outage risk.
We label this cluster \textit{Operational Drag}.\fnmark{interview}

\fntext{interview}{
  The strong rejection of B factors combined with only moderate endorsement of C4 and C6 suggests that this cluster's defining failure mechanism may be narrower or more nuanced than what the 12 survey items can capture.
  In-depth interviews with practitioners who experienced this failure pattern could reveal whether additional operational factors (not represented in the current framework) played a role.
  Such interviews were not feasible as part of the current research project (see Appendix~\ref{sec:survey_admin}) and are left to future work.
}

The smallest cluster ($n = 31$, 7.8\%) is defined by near-universal endorsement of scope creep (B3, 100\%, median 5) and insufficient asset visibility (C2, 93.5\%, median 5), with strong endorsement of complex environment (C1, 87.1\%), unrealistic timeline (B4, 83.9\%), and outage concerns (C6, 83.9\%), all with medians of 5.
This cluster describes projects where the scope expanded beyond what was originally planned in an environment that was difficult to see and understand, compounded by fear of disrupting production systems.
We label this cluster \textit{Scope \& Visibility Trap}.

\begin{table}[htbp]
\centering
\caption{Median Likert responses per cluster ($K = 4$).
Scale: 1 = Strongly Disagree to 5 = Strongly Agree.}
\label{tab:cluster_medians}
\begin{tabular}{lcccc}
\toprule
Item & \multicolumn{4}{c}{Cluster} \\
\cmidrule(lr){2-5}
     & \shortstack{Perfect\\Storm}
     & \shortstack{Diffuse\\Friction}
     & \shortstack{Operational\\Drag}
     & \shortstack{Scope \&\\Visibility Trap} \\
     & ($n$=201) & ($n$=134) & ($n$=34) & ($n$=31) \\
\midrule
B1 & 4 & 4 & 2   & 4   \\
B2 & 4 & 4 & 2   & 2   \\
B3 & 5 & 4 & 3   & 5   \\
B4 & 5 & 4 & 2   & 5   \\
B5 & 4 & 4 & 2   & 2   \\
B6 & 4 & 4 & 2   & 2   \\
\midrule
C1 & 4 & 4 & 2   & 5   \\
C2 & 4 & 4 & 2.5 & 5   \\
C3 & 4 & 4 & 3   & 5   \\
C4 & 4 & 4 & 4   & 4   \\
C5 & 4 & 4 & 2   & 4   \\
C6 & 4 & 4 & 4   & 5   \\
\bottomrule
\end{tabular}
\end{table}

\begin{table}[htbp]
\centering
\caption{Endorsement rates (\% of cluster rating item $\geq$ 4)
per cluster ($K = 4$).}
\label{tab:cluster_endorsement}
\begin{tabular}{lcccc}
\toprule
Item & \multicolumn{4}{c}{Cluster} \\
\cmidrule(lr){2-5}
     & \shortstack{Perfect\\Storm}
     & \shortstack{Diffuse\\Friction}
     & \shortstack{Operational\\Drag}
     & \shortstack{Scope \&\\Visibility Trap} \\
     & ($n$=201) & ($n$=134) & ($n$=34) & ($n$=31) \\
\midrule
B1 & 97.5 & 59.0 & 17.6 & 61.3  \\
B2 & 98.0 & 60.4 &  8.8 & 29.0  \\
B3 & 96.5 & 63.4 & 41.2 & 100.0 \\
B4 & 99.5 & 68.7 & 32.4 & 83.9  \\
B5 & 93.0 & 57.5 & 35.3 & 29.0  \\
B6 & 98.0 & 67.9 & 20.6 & 45.2  \\
\midrule
C1 & 98.5 & 76.1 & 38.2 & 87.1 \\
C2 & 98.0 & 73.9 & 38.2 & 93.5 \\
C3 & 98.5 & 74.6 & 44.1 & 77.4 \\
C4 & 98.5 & 76.1 & 55.9 & 77.4 \\
C5 & 98.5 & 74.6 & 32.4 & 77.4 \\
C6 & 99.0 & 71.6 & 55.9 & 83.9 \\
\bottomrule
\end{tabular}
\end{table}

\subsection{External Validation of Cluster Solution} \label{sec:external_validation}

\begin{figure*}[t] 
  \centering
  \includegraphics[width=1.00\textwidth]{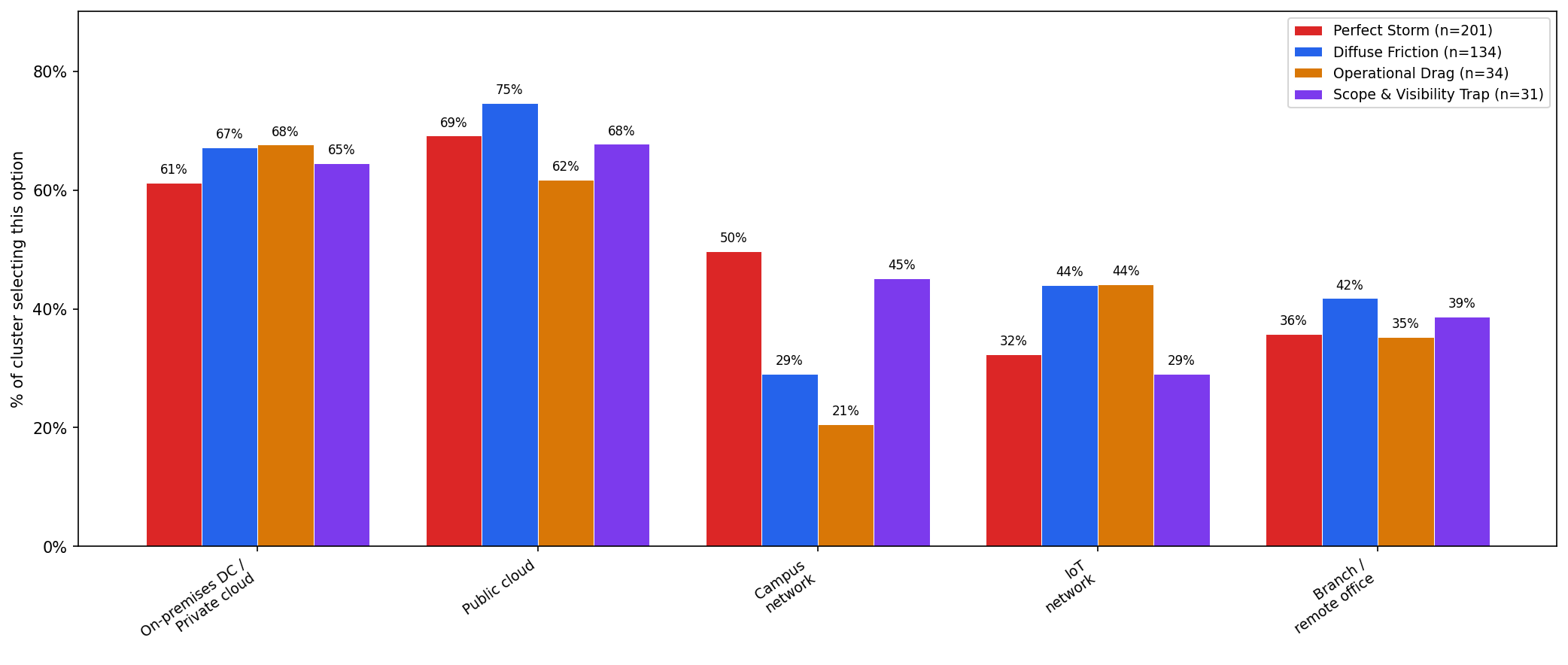}
  \caption{Network Environments in Scope by Cluster (Archetype).}
  \label{fig:environment}
\end{figure*}

\begin{figure}
  \centering
  \includegraphics[width=0.50\textwidth]{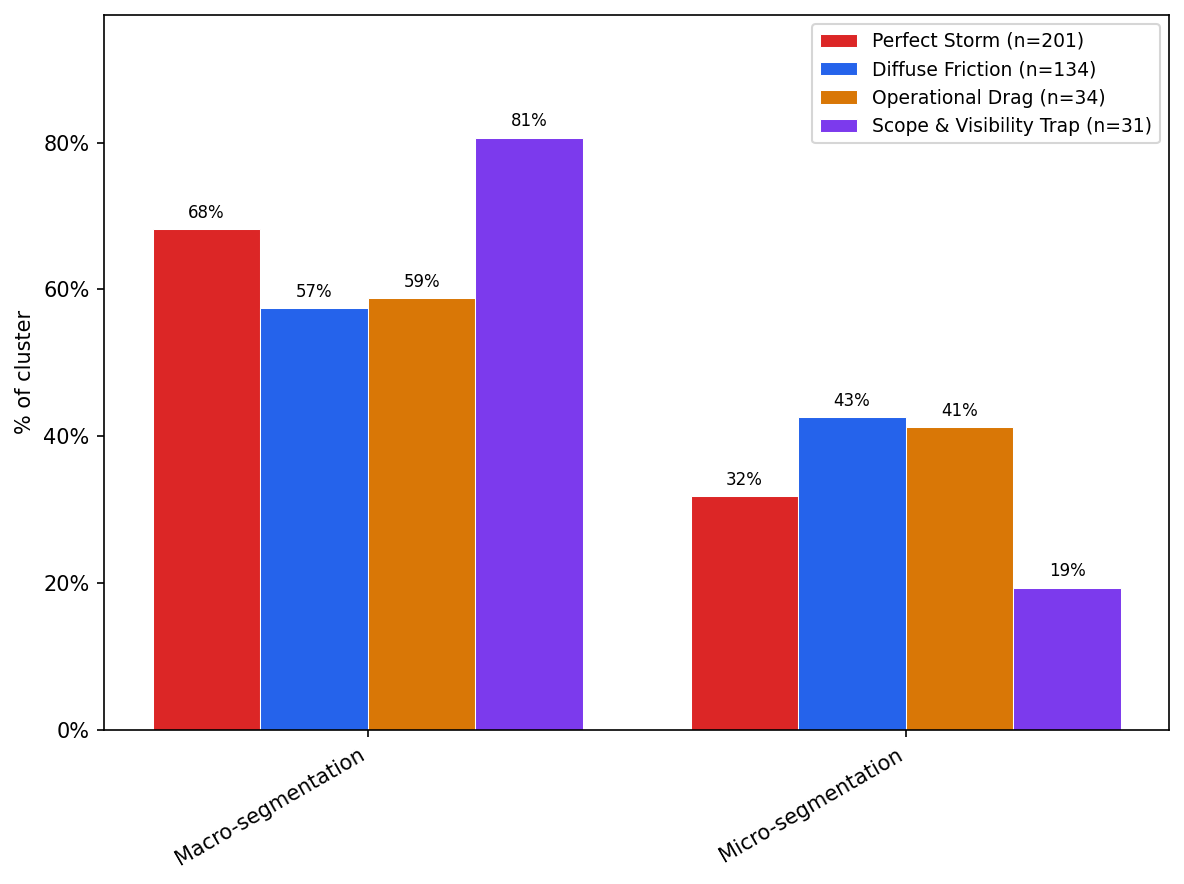}
  \caption{Network Segmentation Model Type by Cluster (Archetype).}
  \label{fig:model}
\end{figure}

\begin{figure*}[t]
  \centering
  \includegraphics[width=1.00\textwidth]{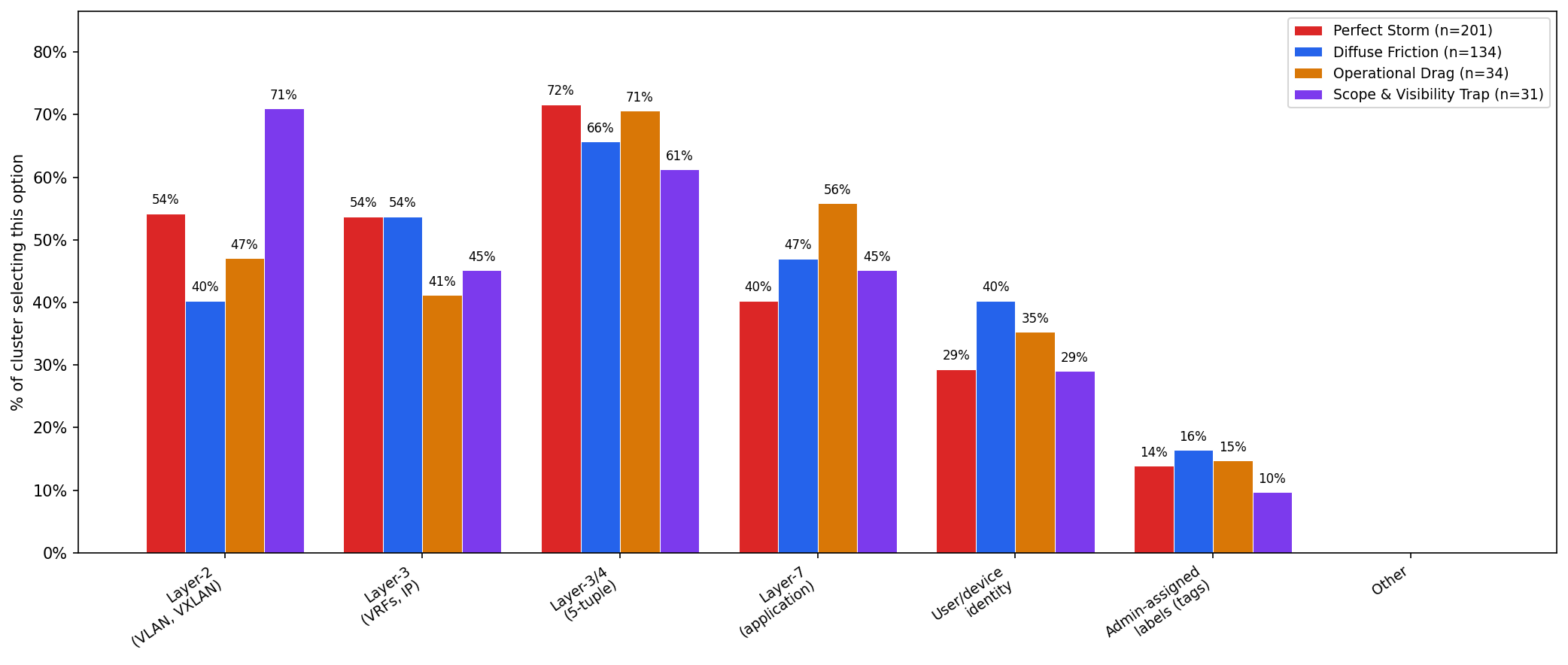}
  \caption{Network Segmentation Approach Type by Cluster (Archetype).}
  \label{fig:approach}
\end{figure*}

\begin{table*}[htbp]
\centering
\caption{External validation variables, short labels, and response options.}
\label{tab:external_variables}
\footnotesize
\begin{tabular}{llll}
\toprule
Variable & Short Label & Type & Response Options \\
\midrule
Q1  & Segmentation model type
    & Single select
    & Macro-segmentation; Micro-segmentation; Don't know \\
Q2  & Segmentation approach
    & Multi-select
    & Layer-2 (VLAN, VXLAN); Layer-3 (VRFs, IP); Layer-3/4 (5-tuple); \\
    & & & Layer-7 (application); User/device identity; \\
    & & & Admin-assigned labels (tags); Other; Don't know \\
Q3  & Network environment
    & Multi-select
    & On-premises DC / private cloud; Public cloud; Campus network; \\
    & & & IoT network; Branch / remote office; Don't know \\
Q4  & Workload type
    & Multi-select
    & Physical (bare metal); Virtualized (ESXi/KVM/Hyper-V); \\
    & & & Container (Docker/K8s); Serverless; Don't know \\
ProfileOrganizationSize
    & Organization size
    & Single select
    & 500--999; 1,000--2,999; 3,000--4,999; 5,000+; Don't know \\
ProfileSuccess
    & Primary failure type
    & Single select
    & Cancelled before implementation; Partially implemented then \\
    & & & paused/cancelled; Fully implemented but rolled back; \\
    & & & Delivered late/over budget (20\%+); \\
    & & & Delivered but failed to meet objectives \\
\bottomrule
\end{tabular}
\end{table*}

\begin{table*}[htbp]
\centering
\caption{Significant and near-significant external associations
with cluster membership ($K = 4$).
Percentages represent the proportion within each cluster.
For multi-select items, percentages reflect the proportion
selecting that option.
``Don't know'' responses excluded per variable.}
\label{tab:external_significant}
\footnotesize
\begin{tabular}{llcccccl}
\toprule
Variable & Category
  & \shortstack{Perfect\\Storm}
  & \shortstack{Diffuse\\Friction}
  & \shortstack{Operational\\Drag}
  & \shortstack{Scope \&\\Visibility Trap}
  & $\chi^2$ & $p$ \\
  & & ($n$=201) & ($n$=134) & ($n$=34) & ($n$=31) & & \\
\midrule
Network environment
  & Campus in scope
  & 49.8 & 29.1 & 20.6 & 45.2
  & 20.27 & $<$0.001 \\
\midrule
Segmentation approach
  & Layer-2 used
  & 54.2 & 40.3 & 47.1 & 71.0
  & 12.04 & 0.007 \\
\midrule
Segmentation model type
  & Macro-segmentation
  & 68.2 & 57.5 & 58.8 & 80.6
  & 8.10 & 0.044 \\
  & Micro-segmentation
  & 31.8 & 42.5 & 41.2 & 19.4
  & & \\
\midrule
Organization size
  & 500--999
  & 23.4 & 26.1 & 11.8 &  6.5
  & 16.87 & 0.051$^\dagger$ \\
  & 1,000--2,999
  & 39.3 & 33.6 & 58.8 & 58.1
  & & \\
  & 3,000--4,999
  & 28.4 & 27.6 & 17.6 & 19.4
  & & \\
  & 5,000+
  &  9.0 & 12.7 & 11.8 & 16.1
  & & \\
\midrule
Network environment
  & IoT in scope
  & 32.3 & 44.0 & 44.1 & 29.0
  & 6.30 & 0.098$^\dagger$ \\
\bottomrule
\multicolumn{8}{l}{\footnotesize
$^\dagger$ Approaching significance;
organization size has 2 of 16 cells with expected count $< 5$.} \\
\end{tabular}
\end{table*}

\begin{table*}[htbp]
\centering
\caption{Non-significant external associations with cluster
membership ($K = 4$).
``Don't know'' responses excluded per variable.}
\label{tab:external_nonsignificant}
\footnotesize
\begin{tabular}{llccl}
\toprule
Variable & Item tested & $\chi^2$ & $p$ & \\
\midrule
Primary failure type
  & Overall
  & 16.68 & 0.162 & $^{\ddagger}$ \\
\midrule
Segmentation approach
  & Layer-3 (VRFs, IP)            &  2.58 & 0.461 & \\
  & Layer-3/4 (5-tuple)           &  2.23 & 0.526 & \\
  & Layer-7 (application)         &  3.56 & 0.313 & \\
  & User/device identity          &  4.66 & 0.199 & \\
  & Admin-assigned labels (tags)  &  1.03 & 0.793 & $^{\dagger}$ \\
\midrule
Network environment
  & On-premises DC / private cloud &  1.47 & 0.689 & \\
  & Public cloud                   &  2.61 & 0.456 & \\
  & Branch / remote office         &  1.33 & 0.721 & \\
\midrule
Workload type
  & Physical (bare metal)          &  3.57 & 0.312 & \\
  & Virtualized (ESXi/KVM/Hyper-V) &  1.89 & 0.595 & \\
  & Container (Docker/K8s)         &  1.10 & 0.777 & \\
  & Serverless                     &  3.53 & 0.317 & \\
\bottomrule
\multicolumn{5}{l}{\footnotesize
$^{\dagger}$ 2 of 8 cells with expected count $< 5$.} \\
\multicolumn{5}{l}{\footnotesize
$^{\ddagger}$ 4 of 20 cells with expected count $< 5$;
test may be underpowered.} \\
\end{tabular}
\end{table*}

The four clusters were derived solely from responses to the general IT project failure items (B1--B6) and the segmentation project failure items (C1--C6).
If the clusters reflect meaningful differences in project characteristics, they should also differ on variables that were not used in the clustering.
To test this, chi-square tests of independence were conducted between cluster membership and six external variables: segmentation model type, segmentation approach, network environment, workload type, organization size, and primary failure type (Table~\ref{tab:external_variables}); because these constitute multiple unadjusted comparisons, the associations are interpreted as exploratory except where noted (see Section~\ref{sec:limitations}).
``Don't know'' responses were excluded on a per-variable basis.
Results are presented in Tables~\ref{tab:external_significant} and~\ref{tab:external_nonsignificant}.

Projects that included campus networks in scope were significantly more likely to fall into the Perfect Storm or Scope \& Visibility Trap clusters ($p < 0.001$).
Campus segmentation introduces architectural complexity beyond data center or cloud environments---spanning building interconnects, wireless infrastructure, and diverse endpoint populations---which is consistent with the broad failure attribution seen in Perfect Storm and the technical overwhelm that characterizes Scope \& Visibility Trap.
By contrast, Diffuse Friction and Operational Drag projects were much less likely to have campus networks in scope (Table~\ref{tab:external_significant}, Figure~\ref{fig:environment}).

The Scope \& Visibility Trap cluster is associated with traditional Layer-2 macro-segmentation ($p = 0.007$ for Layer-2 adoption; $p = 0.044$ for macro-segmentation model type).
This cluster has the highest rates of both macro-segmentation and Layer-2 adoption of any group.
Layer-2 macro-segmentation using VLANs and VXLANs depends heavily on accurate asset visibility to define effective network zones, which aligns directly with this cluster's profile: universal endorsement of scope creep and near-universal endorsement of insufficient asset visibility.
The remaining three clusters are more evenly split between macro and micro approaches (see Figures~\ref{fig:model},~\ref{fig:approach}).

Two additional variables approach but do not reach statistical significance.
The Operational Drag and Scope \& Visibility Trap clusters (the two smaller, selective-attribution groups) concentrate heavily in the 1,000--2,999 employee range, while Perfect Storm and Diffuse Friction are more evenly distributed across size categories ($p = 0.051$; Table~\ref{tab:external_significant}).
This result should be interpreted with caution due to low expected cell counts.\fnmark{chi-square}
IoT network scope ($p = 0.098$) shows a different grouping: Diffuse Friction and Operational Drag have notably higher IoT involvement than Perfect Storm and Scope \& Visibility Trap.
Both findings warrant further investigation with a larger sample.

\fntext{chi-square}{
  The chi-square test compares observed cell counts to the counts that would be expected if cluster membership and the variable were unrelated~\cite{agresti2007introduction}.
  When a small cluster is crossed with a variable that has many categories, some cells have expected counts below 5, at which point the chi-square approximation becomes unreliable and the resulting p-value may be inaccurate.
}

No significant differences between clusters were detected for the remaining external variables in the survey (Table~\ref{tab:external_nonsignificant}).
Primary failure type, most segmentation approaches (Layer-3, Layer-3/4, Layer-7, identity-based, and tag-based), most network environments (on-premises, public cloud, and branch/remote office), and all four workload types are distributed similarly across the four clusters.
This suggests that the clusters are not simply proxies for workload composition or failure type, although the small size of two clusters limits the power of these tests.

Taken together, the significant associations involve variables that describe \textit{how} segmentation was attempted: macro- vs. micro-segmentation, the specific segmentation approach (Layer-2 vs. Layer-3 etc.), and the specific network types in scope.
Variables describing \textit{what} workloads were involved or \textit{what} failure types occurred show no significant association with cluster membership.

\section{Practitioner-proposed Remedies} \label{sec:what}

The previous section identified four distinct failure archetypes, each characterized by a different pattern of attribution across B and C factors.
A natural follow-up question is whether practitioners who experience different types of failure also propose different remedies.
If so, the fix strategy should be tailored to the archetype.
If not, the disconnect between diagnosis and prescription is itself a finding with practical implications.

\begin{figure}
  \centering
  \includegraphics[width=0.50\textwidth]{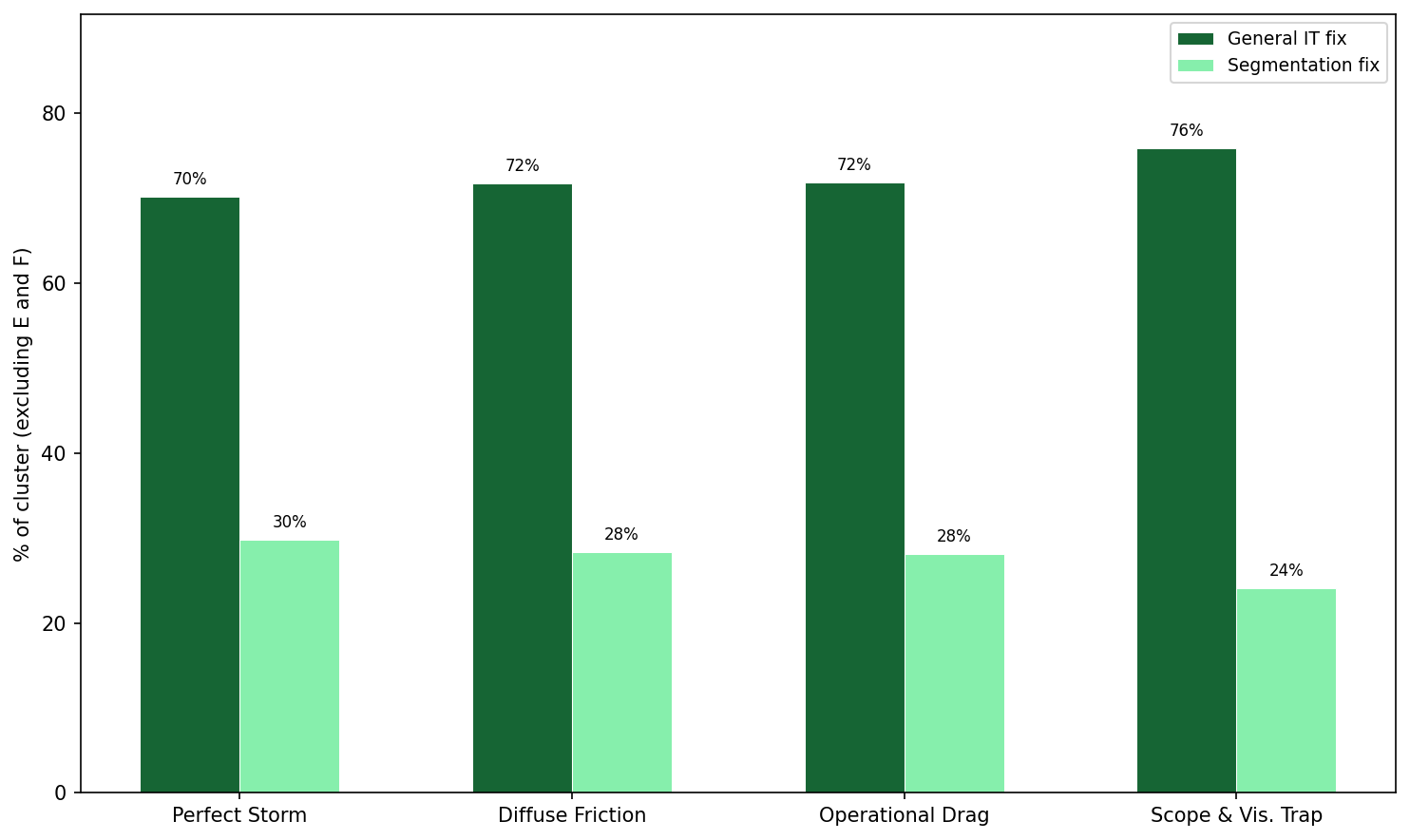}
  \caption{Proposed Fix Category by Failure Archetype.}
  \label{fig:fix_category}
\end{figure}

To investigate this, we analyze responses to Q8, which asked: \textit{``In your view, what is the single most important change that you would implement if you could do this segmentation project again?''}
Each response was coded against a codebook aligned with the B1--B6 and C1--C6 failure framework (see Appendix~\ref{sec:coding}) and collapsed into two categories: general IT project management fixes (B) and segmentation-specific fixes (C)~\cite{hsieh2005three}.
A small number of responses (6.0\%) that were either emergent or non-actionable were excluded.
Table~\ref{tab:fix_by_cluster} (Figure~\ref{fig:fix_category}) presents the distribution of proposed fixes across the four failure archetypes.

All four archetypes propose fixes in virtually the same ratio: approximately 70\% B fixes and 30\% C fixes.
This ratio holds regardless of whether the cluster attributed failure broadly (Perfect Storm and Diffuse Friction) or selectively to C factors (Operational Drag and Scope \& Visibility Trap).
The chi-square test confirms no significant association between failure archetype and proposed fix category ($p = 0.94$).
Practitioners who experience fundamentally different types of failure converge on the same balance of remedies (see Appendix~\ref{sec:sample} for sample responses).

This convergence is particularly notable for the two selective-attribution clusters.
Operational Drag respondents endorsed B factors at low rates (Table~\ref{tab:cluster_endorsement}), identifying excessive manual effort in policy maintenance and outage risk as their primary barriers.
Yet 71.9\% propose a general IT fix.
Similarly, Scope \& Visibility Trap respondents attributed failure primarily to segmentation-specific challenges such as insufficient asset visibility and environmental complexity, alongside scope creep (B3) and unrealistic timeline (B4); yet 75.9\% propose a general IT fix.

Two complementary explanations may account for this pattern.
First, general project management shortcomings may shape how practitioners experience a project overall.
Even when the proximate cause of failure is segmentation-specific, working within a project that struggled with some aspects of project management may leave respondents feeling that the project was fundamentally mishandled, and their proposed fix reflects that broader sentiment.
Second, general project management failures may be the upstream root cause.
If the project had been scoped realistically, resourced adequately, and governed effectively, the segmentation-specific challenges might have been discovered earlier and resolved before they became fatal.
Under this reading, respondents are not contradicting their diagnosis: they are looking past the proximate technical barrier to the organizational conditions that allowed it to emerge.
Both explanations lead to actionable guidance.
Projects must be run well at a minimum: participants need to believe the project is worthwhile and has a realistic chance of success.
However, when a segmentation-specific failure is identified, it is that failure that needs direct attention.
Resorting to general IT project management fixes alone will not make a segmentation-specific problem go away; both the upstream conditions and the proximate technical barrier may need to be addressed simultaneously.

Future research could investigate whether the observed preference for B fixes over C fixes reflects a default toward more familiar remedies, a genuine causal belief, or something else altogether.

\begin{table}[htbp]
\centering
\caption{Proposed fix category by failure archetype (\% of cluster).
Responses coded as emergent or null (6.0\% of total) are excluded.
$\chi^2 = 0.42$, $df = 3$, $p = 0.94$.}
\label{tab:fix_by_cluster}
\begin{tabular}{lccc}
\toprule
         &     & General IT & Segmentation \\
Cluster  & $n$ & Fix (\%)   & Fix (\%) \\
\midrule
Perfect Storm           & 188 & 70.2 & 29.8 \\
Diffuse Friction         & 127 & 71.7 & 28.3 \\
Operational Drag    &  32 & 71.9 & 28.1 \\
Scope \& Vis.\ Trap    &  29 & 75.9 & 24.1 \\
\bottomrule
\end{tabular}
\end{table}

\section {Limitations} \label{sec:limitations}

The findings in this study are based on practitioner survey responses and are therefore subject to common limitations of survey-based research.
Survey respondents may not fully represent all organizations or roles involved in segmentation projects, and practitioners with stronger experiences or opinions may be more likely to respond.
The survey captures practitioner perceptions using a limited number of items, which necessarily provide a simplified view of complex organizational and technical issues, and answering additional questions at finer granularity would require designing and fielding a separate survey.
Responses also reflect individual perspectives and retrospective judgment, which may be influenced by role, experience, or organizational context, and each project is described by a single respondent rather than corroborated by a second team member or an independent project artifact.
The anonymized panel through which the survey was fielded (Appendix~\ref{sec:survey_admin}) means that no respondent-identifying information reached the author, so validating individual project descriptions against additional informants or artifacts was not possible by design.
Multi-informant or artifact-based validation is a worthwhile direction for future work, though it would require a fundamentally different study design that trades away the respondent anonymity on which panel-based surveys of this kind depend.
In addition, because segmentation practices and constraints vary across industries, organization sizes, and technology environments, the identified patterns should be interpreted as broadly indicative rather than universally applicable to all segmentation deployments (also see Appendix~\ref{sec:generalizability}).

The clustering results further depend on how failure factors were defined and measured.
Failure factors were represented using twelve Likert-scale survey questions based on prior research and industry reports.
This focused set of questions supports interpretability, but different question formulations or additional failure factors could lead to different clustering results.
Accordingly, the identified clusters should be viewed as patterns within the chosen set of failure factors rather than as a complete or authoritative explanation of why segmentation projects fail.

The two smaller archetypes (Operational Drag and Scope \& Visibility Trap) contain 34 and 31 respondents respectively.
Both clusters fall within the 5--8\% range suggested for adequate class size~\cite{nylund2018ten}, but their small absolute counts limit the precision of within-cluster statistics and the power of external validation tests.
A larger sample might reveal additional structure within these groups or sharpen the borderline findings for organization size and IoT network scope.

As noted where the tests are introduced (Section~\ref{sec:external_validation}), the external validation analysis involves multiple chi-square comparisons that were not adjusted for multiple testing.
Consequently, some reported associations may be false positives, particularly those with marginal significance.
The campus-in-scope result is the exception: it is the strongest association and is treated as robust, whereas the remaining associations are interpreted as exploratory.

The analysis of Q8 responses also carries interpretive limitations.
Each free-form response was assigned a code reflecting the researchers' reading of the respondents' intent, so ambiguous responses may have been interpreted differently than the respondent meant.
In addition, each response received exactly one code; responses that proposed multiple fixes of different types were reduced to the most central one, which may undercount secondary fix types.
Finally, Q8 asked respondents what they would change, not what actually caused the failure.
Because people tend to name changes they feel able to make, the shared preference for general project management fixes may reflect what respondents felt they could control rather than what would have worked best.

\section{Conclusion} \label{sec:conclusion}

The 2025 National Academies Cyber Hard Problems report identifies the lack of an empirical basis for security decisions as a fundamental challenge, observing that most established cybersecurity best practices rest on common sense and received wisdom rather than rigorous evidence \cite{national2025cyber}.
This paper responds to that call by providing what is, to the best of our knowledge, the first systematic empirical survey of why network segmentation projects fail.

We developed a survey grounded in a two-part failure framework that distinguishes general IT project failure factors from segmentation-specific failure factors.
Both framework components were measured using six survey items each.
The resulting survey was administered to 400 network security practitioners in the U.S.\

Latent Class Analysis of the survey responses identified four distinct failure archetypes.
\textit{Perfect Storm} (50.2\% of respondents) describes projects in which general IT project management failures and segmentation-specific technical challenges occurred simultaneously and pervasively.
\textit{Diffuse Friction} (33.5\%) describes projects that stalled under the cumulative weight of broad, moderate challenges rather than failing on any single front.
\textit{Operational Drag} (8.5\%) describes projects where there was adequate goal clarity and executive sponsorship, but the operational burden of policy creation and maintenance proved unsustainable.
\textit{Scope \& Visibility Trap} (7.8\%) describes projects defeated by scope changes, unrealistic timeline, and the technical difficulty of segmentation in a complex environment with poor asset visibility and low disruption tolerance.

The archetypes are not merely statistical groupings: they correspond to real differences in how segmentation was attempted.
Projects involving campus networks and traditional Layer-2 macro-segmentation are more likely to fall into the Perfect Storm or Scope \& Visibility Trap archetypes.
Practitioners planning segmentation initiatives that span campus environments or rely on Layer-2 approaches should anticipate a higher risk of broad or technically intense failure and plan accordingly.
By contrast, the archetypes do not significantly differ by workload type, suggesting that the failure patterns arise regardless of whether the environment runs bare metal, virtualized, containerized, or serverless workloads.

When asked what single change they would make, respondents across all four archetypes proposed general IT project management fixes over segmentation-specific fixes in the same roughly 70/30 ratio, even when their failure attribution was predominantly segmentation-specific.

These findings carry direct implications for practitioners managing segmentation initiatives.
The existence of four distinct failure archetypes suggests that a one-size-fits-all approach to segmentation project recovery is insufficient: a Perfect Storm project requires broad organizational remediation, while a Scope \& Visibility Trap project needs targeted investment in asset discovery and environmental scoping.
At the same time, the convergence on general IT project management fixes across all archetypes reveals a gap between how practitioners diagnose failure and how they propose to address it.
Organizations should treat strong project governance as a necessary foundation, but when segmentation-specific barriers are identified---whether excessive policy burden, poor asset visibility, or environmental complexity---those barriers require segmentation-specific interventions rather than a retreat to general project management practices.

\section*{Acknowledgments}
This work was supported by Cisco Systems Inc.

\appendices

\section{Survey Administration and Ethics} \label{sec:survey_admin}

The survey was administered on our behalf by Vanson Bourne,\fnmark{vb} an independent research firm that maintains a proprietary panel of IT and security professionals and recruits respondents from it.
In total, 4,921 panelists were exposed to the survey.
Of this initial pool, 3,050 individuals were screened out during the qualification phase for failing to meet the eligibility criteria.
These criteria included involvement in a network segmentation project within the previous 24 months and employment at an organization meeting the required employee count (500) at the time of the project.
A further 1,271 respondents who passed the screening phase were eliminated following stringent data quality checks.
These exclusions were based on behaviors such as ``speeding'' through the survey or providing a high frequency of ``Don't know'' responses.
Another 182 qualified respondents dropped out of the survey during the profiling section of the survey.
Furthermore, 18 respondents dropped out in the main part of the survey; some of these were terminated once the target sample size of 400 was achieved and the survey was closed.

\begin{figure*}
  \centering
  \includegraphics[width=0.75\textwidth]{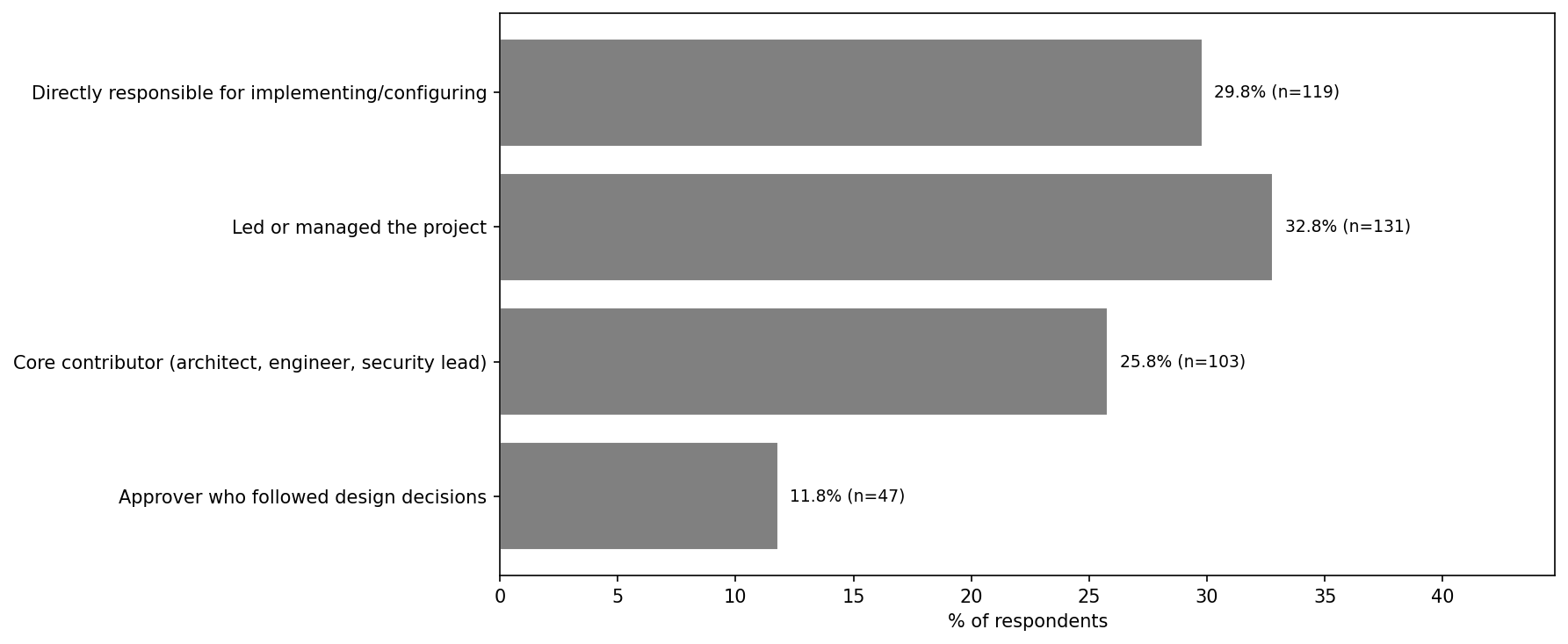}
  \caption{Role Distribution of Respondents on Failed Segmentation Project.}
  \label{fig:sample_role}
\end{figure*}

\begin{figure*}
  \centering
  \includegraphics[width=0.75\textwidth]{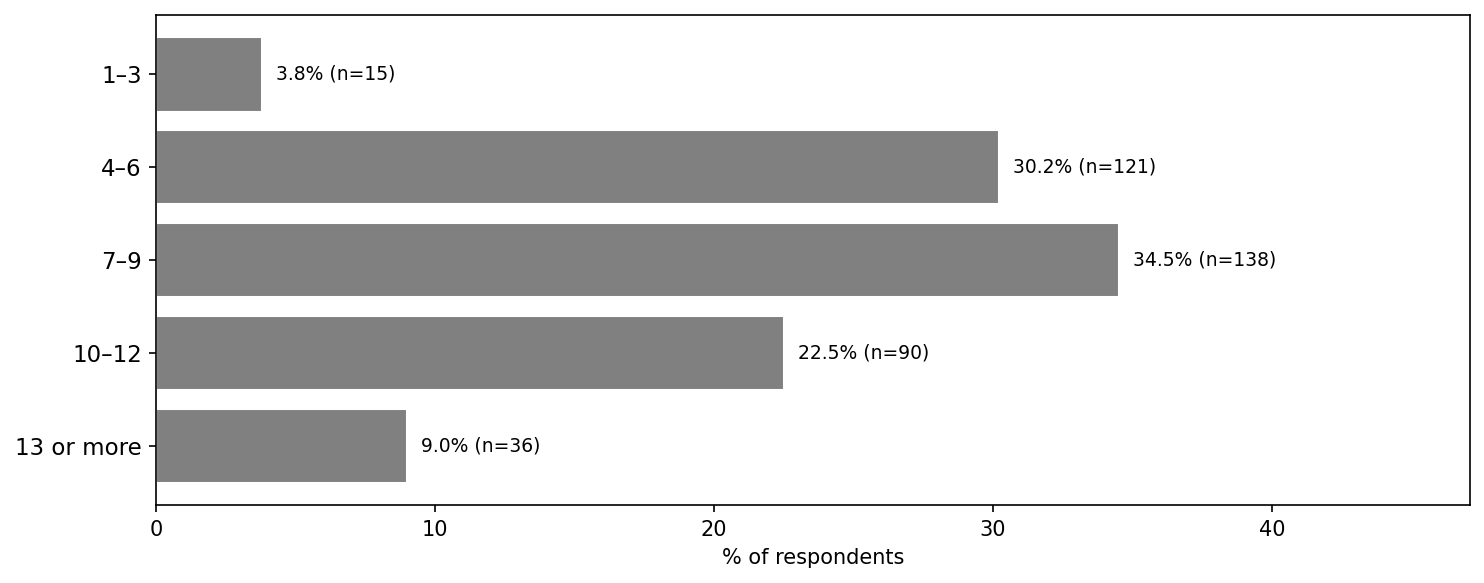}
  \caption{Segmentation Projects Undertaken by Respondents over Career.}
  \label{fig:sample_career_projects}
\end{figure*}

The final sample comprises 400 completed responses from U.S.-based\ network security practitioners.
Each respondent reported on a single failed segmentation project occurring within the previous 24 months at an organization with 500 or more employees.
The median completion time for a successful response was 16~minutes and 58~seconds.
For information on the project role and career experience of the 400 respondents see Figures~\ref{fig:sample_role},~\ref{fig:sample_career_projects}.

Although the study was commissioned by Cisco Systems Inc., it was designed and executed through the research firm's panel-based model, which is built around respondent privacy.
The firm guarantees anonymity to its panelists, and none of the responses made available to the author contained any information identifying individual respondents or their employers.
Panelists selected to participate were compensated by the firm in accordance with its standard panel practices, and participation was voluntary, with consent obtained through the firm's standard enrollment and survey consent procedures.
Because the author worked only with de-identified responses and had no interaction with human subjects, the study did not meet the criteria requiring review by an institutional review board.

\fntext{vb}{
  \url{https://www.vansonbourne.com/}
}

\section{Scope and Generalizability} \label{sec:generalizability}

The design of this study reflects a series of deliberate trade-offs made to obtain high-quality practitioner data within practical constraints.
Fielding a survey of this kind through a professional panel is expensive, and cost scales with sample size, screening stringency, and the breadth of populations targeted.
Several scope decisions follow directly from this reality and bound the populations to which the findings apply.

First, the study was deliberately restricted to U.S.-based\ organizations with 500 or more employees.
Limiting the geography and the organization size range kept both survey cost and project complexity contained, at the expense of coverage.
The findings should therefore be read as applying to segmentation projects at reasonably large U.S.\ enterprises, and not necessarily to smaller organizations, to non-U.S.\ contexts, or to environments with materially different regulatory and operational conditions.

Second, the study reports results only for the 400 respondents who passed screening and data quality checks (Appendix~\ref{sec:survey_admin}).
Over 85\% of the initial pool was removed, the large majority through eligibility screening rather than quality exclusions.
This funnel is a feature of the design rather than sample attrition: it selects for practitioners with genuine, recent, first-hand experience of a failed segmentation project at a qualifying organization.
The resulting sample is well matched to the study's question but is not a random sample of the broader practitioner population, and the reported patterns should be interpreted accordingly.

Third, the achievable sample size was too small to support reliable between-group comparisons along dimensions such as industry sector (for example, healthcare versus financial services).
Such comparisons partition the sample into progressively smaller cells, and the two smaller archetypes already contain only 34 and 31 respondents.
Questions of whether, say, heavily regulated sectors experience distinct failure patterns are important, but answering them credibly would require a substantially larger and differently stratified sample.
We therefore leave sector-level (and other fine-grained) comparisons to future work.

Finally, the study focuses on failure factors rather than attempting to measure success factors or to include a comparison group of successful projects within the same instrument.
Respondent response time is a binding constraint: the median completion time was already close to seventeen minutes (Appendix~\ref{sec:survey_admin}), and expanding the instrument to cover both failure and success dimensions in comparable depth would have lengthened it beyond what sustains reliable, high-quality responses.
Given this constraint, we chose depth on failure over breadth across outcomes, and considerable diligence went into grounding the twelve failure factors in the prior academic and industry literature (Sections~\ref{sec:it_failure_framework} and~\ref{sec:seg_failure_framework}).
A consequence is that the factors identified here characterize failed projects and cannot, on their own, establish that these factors are absent from successful ones; disentangling failure-specific from generally present factors would require a matched comparison group and is a natural direction for subsequent research.

\section{Additional Evidence for Four Classes} \label{sec:stability}

To assess whether the four-class solution is robust to sampling variability, the LCA was refitted on 200 bootstrap samples (resampled with replacement) for $K = 3$, $K = 4$, and $K = 5$.
Within each bootstrap sample (and for each $K$), respondents were assigned to the class with the highest posterior probability to obtain hard class labels for comparison.
Class assignments from each bootstrap sample were compared to the corresponding full-sample assignments using the Adjusted Rand Index (ARI), which measures agreement between two sets of class labels on a scale from 0 (random) to 1 (perfect)~\cite{hubert1985comparing}.
Table~\ref{tab:bootstrap_stability} presents the results.
The $K = 4$ solution has comparable stability to $K = 3$, with a higher mean and median ARI and lower variability between bootstrap runs.
At $K = 5$, stability drops noticeably, with no bootstrap samples reaching the 0.80 threshold for good agreement.
The moderate absolute ARI values at $K = 4$ are consistent with expectations for a model containing two smaller classes of 31 and 34 respondents; bootstrap samples that underrepresent one of these groups will produce a less precise recovery of the full-sample solution.
The stability analysis therefore establishes an upper bound ($K < 5$), since $K = 5$ fragments the data beyond what can be reliably recovered.

To establish a lower bound, the Bootstrap Likelihood Ratio Test (BLRT) was used to test whether each successive class provides a statistically significant improvement in model fit over the previous solution~\cite{nylund2007deciding,mclachlan2000finite}.
Table~\ref{tab:blrt} presents the results.
All four tests are significant ($p < 0.01$), indicating that each additional class from $K = 2$ through $K = 5$ captures structure not present in the simpler model.
In particular, the $K = 4$ vs. $K = 3$ test is significant with a likelihood ratio statistic of 220.4, confirming that the fourth class represents a genuine improvement over the three-class solution rather than an artifact of model flexibility.
The BLRT therefore establishes a lower bound ($K > 3$), since stopping at $K = 3$ would discard meaningful structure that the data supports.

Taken together, the two tests bracket the solution.
The bootstrap stability analysis rules out $K = 5$ (the fifth class cannot be reliably recovered), while the BLRT rules out $K = 3$ (the fourth class is statistically justified).
$K = 4$ is the only solution that satisfies both criteria.

\begin{table}[htbp]
\centering
\caption{Bootstrap stability of LCA solutions ($K = 3$ to $K = 5$).
Adjusted Rand Index (ARI) computed over 200 shared bootstrap samples.}
\label{tab:bootstrap_stability}
\footnotesize
\begin{tabular}{lccccc}
\toprule
$K$ & Mean ARI & SD & Median ARI & Q25--Q75 & \% $\geq$ 0.80 \\
\midrule
3 & 0.637 & 0.155 & 0.651 & 0.559--0.749 & 15.0 \\
4 & 0.655 & 0.117 & 0.662 & 0.571--0.751 &  9.5 \\
5 & 0.592 & 0.104 & 0.608 & 0.527--0.662 &  0   \\
\bottomrule
\end{tabular}
\end{table}

\begin{table}[htbp]
\centering
\caption{Bootstrap Likelihood Ratio Test (BLRT) results.
100 parametric bootstrap samples per test.}
\label{tab:blrt}
\footnotesize
\begin{tabular}{lcccc}
\toprule
Test & LL$_{K-1}$ & LL$_{K}$ & LRT & $p$ \\
\midrule
$K=2$ vs. $K=1$ & $-$5842.4 & $-$5449.3 & 786.2 & $<$0.01 \\
$K=3$ vs. $K=2$ & $-$5449.3 & $-$5280.3 & 338.1 & $<$0.01 \\
$K=4$ vs. $K=3$ & $-$5280.3 & $-$5170.1 & 220.4 & $<$0.01 \\
$K=5$ vs. $K=4$ & $-$5170.1 & $-$5099.1 & 142.0 & $<$0.01 \\
\bottomrule
\end{tabular}
\end{table}

\section{Response-Style Check} \label{sec:response_style}

Because the Perfect Storm archetype endorses nearly every failure factor (Table~\ref{tab:cluster_endorsement}), a natural concern is that it reflects acquiescence bias, that is, a group of respondents who agreed with almost everything regardless of content, rather than a substantive failure pattern.
Acquiescence is a property of individual respondents rather than of clusters, so the question is specifically whether the respondents grouped into Perfect Storm are yea-sayers.
Several lines of evidence bear on this question.

First, the survey was professionally administered and the respondent pool was screened with care (Appendix~\ref{sec:survey_admin}).
Eligibility screening restricted participation to practitioners with direct, recent experience of a failed segmentation project, and stringent data quality checks removed respondents who sped through the survey or gave a high frequency of ``Don't know'' answers.
These steps do not eliminate acquiescence, but they remove its most common crude sources and lower the prior probability that a cluster is a response-style artifact before any cluster-specific analysis is considered.

Second, and most directly, Perfect Storm membership is associated with external project characteristics that were not used in clustering: campus network scope and Layer-2 adoption (Section~\ref{sec:external_validation}).
Acquiescent responding is content-independent by definition and therefore cannot produce an association between a respondent's cluster and objective attributes of the project they reported.
The presence of such associations is difficult to reconcile with the view that Perfect Storm is simply a group of undifferentiated agreers.

Third, pure straight-lining is rare within the cluster.
Only 12 of the 201 Perfect Storm respondents (6\%) gave the identical response to all twelve items, and every one of these chose 4 (``Slightly agree'') rather than the scale maximum of 5.
A respondent agreeing reflexively to maximize agreement would be expected to gravitate toward the ceiling, so even this small group does not fit the classic yea-saying profile.
The remaining 94\% of the cluster differentiated between items, which is inconsistent with a cluster composed of respondents who agreed uniformly.

Fourth, overall agreement level is not the dimension that separates the archetypes.
Diffuse Friction and Scope \& Visibility Trap have nearly identical overall agreement levels yet form clearly distinct archetypes with different response patterns (Table~\ref{tab:response_style}).
This indicates that the latent class analysis is grouping respondents by the shape of their attribution across factors rather than by how much they agreed overall, which is what a response-style artifact would instead produce.

Finally, we note that acquiescence and genuine broad agreement cannot be distinguished with complete certainty, because a respondent who sincerely believes that every factor contributed produces a response vector nearly identical to that of a yea-sayer.
The paragraphs above therefore establish the balance of evidence rather than proof: professional administration lowers the prior, the external associations and the scarcity of straight-lining argue against a response-style explanation, and the separation of the archetypes on response pattern rather than agreement level indicates that Perfect Storm reflects a substantive failure pattern.

\begin{table}[htbp]
\centering
\caption{Response-style indices by archetype.
The agreement level is each respondent's average across all twelve
items (scale 1--5); the reported SD is the between-respondent spread
of this average within the archetype.
Within-respondent dispersion is the average, across respondents in
the archetype, of how much a single respondent's twelve responses
vary about their own mean.
The B$-$C contrast is the mean B-factor response minus the mean
C-factor response, where negative values favor
segmentation-specific factors.}
\label{tab:response_style}
\footnotesize
\begin{tabular}{lcccc}
\toprule
Archetype & $n$ & \shortstack{Agreement \\ (SD)} & \shortstack{Within-Resp. \\ Dispersion} & \shortstack{B$-$C \\ Contrast} \\ 
\midrule
Perfect Storm      & 201 & 4.4 (0.2) & 0.50 & $-0.01$ \\
Diffuse Friction   & 134 & 3.8 (0.3) & 0.78 & $-0.22$ \\
Operational Drag   &  34 & 2.7 (0.5) & 1.01 & $-0.48$ \\
Scope \& Vis. Trap &  31 & 3.7 (0.5) & 1.46 & $-0.73$ \\
\bottomrule
\end{tabular}
\end{table}

\section{Free-form Text Coding Methodology} \label{sec:coding}

\begin{table*}[htbp]
\centering
\caption{Codebook for free-form response coding.}
\label{tab:codebook}
\footnotesize
\begin{tabularx}{\textwidth}{l >{\hsize=0.4\hsize}X >{\hsize=1.0\hsize}X >{\hsize=1.6\hsize}X}
\toprule
Code & Name & Definition & Examples \\
\midrule
B1 & Clear Goals
   & Define goals clearly and consistently at the outset.
   & If I could change one thing then I would define success more clearly;
     We don't have a segmentation approach and it would be good to have one;
     More clarity on what success looks like \\
B2 & Executive Sponsorship
   & Obtain sufficient senior leadership sponsorship and decision-making support.
   & Clear ownership for segmentation;
     Teams don't agree on priorities;
     Better alignment and leadership \\
B3 & Scope Management
   & Prevent scope creep and changing requirements.
   & If I had the opportunity to change one thing about segmentation it would be the scope;
     Understand what is actually in scope;
     Requirements keep changing \\
B4 & Timeline and Resources
   & Develop realistic project timeline given available resources, including workforce skills and training.
   & We need a larger budget for this critical work;
     Less worry about budget pressures;
     We need more budgeting to help implement a new approach;
     Educating and training people on the benefits and pathway forward. A workshop to be held with an SME to provide information and take questions would help;
     Having more knowledgeable employees to help with decision making process \\
B5 & Project Execution
   & Identify, track and effectively address issues and project risks during execution including phasing the project as a risk reduction measure.
   & We sometimes wait until change is needed rather than proactively planning;
     Planning is good but execution is weak;
     No clear roadmap \\
B6 & Coordination and Communication
   & Coordinate and communicate effectively between stakeholders.
   & I would like different departments to have better understanding;
     More collaboration between teams, then more execution;
     Have full buy in from all areas;
     Creating a rapid customer feedback mechanism enables timely feedback into product and service improvement;
     Increase the focus on customer feedback \\
\midrule
C1 & Architecture and Environment
   & Simplify the environment (e.g., hybrid cloud, containers, legacy systems). Restrict segmentation projects to specific environments. Environmental heterogeneity independent of tooling.
   & Adapt to dynamic environments such as clouds and containers with microsegmented platforms;
     Environment is too complex;
     We need more scalability and create more environments;
     The most important thing is to enhance support for legacy systems and strengthen perimeter protection \\
C2 & Asset Visibility
   & Obtain sufficient visibility into assets to design effective segmentation policies.
   & It needs to be more focused on protecting critical assets to reduce threats, risks, and vulnerabilities;
     Better visibility into assets;
     We don't know all assets in the environment \\
C3 & Communication Flows
   & Identify legitimate communication flows between systems effectively.
   & Shift towards more dynamic and granular behavior-based segmentation;
     I would have my organization use more dynamic approaches;
     Segmented based on the flow of data generation, storage, transmission, use, and destruction;
     Know which systems talk to each other \\
C4 & Policy Lifecycle and Operations
   & Automate creation and maintenance of segmentation policies. Use Artificial Intelligence (AI) to generate or modify policies dynamically, reducing manual work. Standardization of policies.
   & Implementation of AI to improve efficiency and accuracy;
     Make it more automated, policies and rules are manual;
     Tie firewall policy for macro segmentation and host based micro segmentation policies together;
     I'd streamline policy management by moving to a more unified platform, so teams spend less time coordinating across different tools and enforcement points;
     Clearer policy rules across the whole organization \\
C5 & Tooling Maturity
   & Obtain access to mature segmentation tools. Tooling capability, maturity, integration, or 3rd party services regardless of environment.
   & Segmentation tools are immature;
     Better technology implementation and risk mediation;
     We lack tools that support segmentation;
     It would be better integration testing. I think running more parallel with another production environment would be more helpful \\
C6 & Risk Appetite
   & Be willing to take some risk of application outages or business disruption.
   & Enhanced segmentation planning that includes risk;
     Not willing to accept outages;
     Risk tradeoffs are not discussed \\
\midrule
E  & Emergent Fix
   & An emergent category. Use only when the response is clear and specific but does not map to one of B1--B6 or C1--C6 (as the most important fix). If a response includes many fixes but none dominate clearly, use E.
   & Easier navigation;
     Fast forward service delivery;
     Let the new professionals demonstrate the new tools;
     It need [sic] to be more unified \\
F  & Null / No Change
   & Satisfied, unsure or lacks the knowledge to answer. Non-actionable responses. Statements without a proposed change. Any response that is vague, underspecified, or unclear in what it is referring to.
   & nothing [sic], it has a good approach;
     More comprehensive;
     I am very satisfied with the current situation \\
\bottomrule
\end{tabularx}
\end{table*}

\begin{table}[htbp]
\centering
\caption{Distribution of codes across 400 respondents.}
\label{tab:code_distribution}
\footnotesize
\begin{tabular}{llcc}
\toprule
Code & Code Name & $n$ & \% \\
\midrule
B1 & Clear Goals                    &  34 &  8.5 \\
B2 & Executive Sponsorship          &  21 &  5.2 \\
B3 & Scope Management               &  11 &  2.8 \\
B4 & Timeline and Resources         &  45 & 11.2 \\
B5 & Project Execution              & 104 & 26.0 \\
B6 & Coordination and Communication &  53 & 13.2 \\
\midrule
   & \textit{Total B}               & 268 & 67.0 \\
\midrule
C1 & Architecture and Environment   &   4 &  1.0 \\
C2 & Asset Visibility               &  30 &  7.5 \\
C3 & Communication Flows            &  28 &  7.0 \\
C4 & Policy Lifecycle and Operations &  10 &  2.5 \\
C5 & Tooling Maturity               &  35 &  8.8 \\
C6 & Risk Appetite                  &   1 &  0.2 \\
\midrule
   & \textit{Total C}               & 108 & 27.0 \\
\midrule
E  & Emergent Fix                   &  15 &  3.8 \\
F  & Null / No Change               &   9 &  2.2 \\
\midrule
   & \textbf{Total}                 & 400 & 100.0 \\
\bottomrule
\end{tabular}
\end{table}

This appendix describes the methodology used to code free-form responses to Q8, which asked respondents to identify the single most important change they would implement if they could repeat their segmentation project.
The goal was to assign each response a single code representing the most central fix proposed by the respondent.

A 14-code codebook was developed, aligned with the failure framework used in the survey (Tables~\ref{tab:survey_items} and~\ref{tab:segmentation_survey_items})~\cite{hsieh2005three}.
Twelve codes (B1--B6, C1--C6) map directly to the general IT project failure factors and segmentation project failure factors measured by the Likert items.
An additional code captures responses that are concrete but do not map to any of the 12 factors (E, Emergent Fix).
A final code captures responses that are vague, non-actionable, or express satisfaction with the current approach (F, Null/No Change).
Table~\ref{tab:codebook} presents the full codebook with definitions and examples.
Because the codebook describes proposed remedies rather than failure factors, some definitions are broader than the corresponding survey items: for example, the B5 codebook definition includes phasing as a specific risk reduction technique, whereas the B5 survey item measures the general failure to manage risks during execution.
The examples in the codebook were developed after reviewing the free-text responses from another survey and those from the pilot phase of the survey in this paper (the 15 pilot responses are not part of the 400 survey responses analyzed in this paper).

Each response was assigned exactly one code, representing the most central fix in the respondent's answer.
When a response included multiple fixes, the fix receiving the greatest emphasis or most specific elaboration was selected.
When multiple fixes were present but none dominated clearly, E was assigned if the content was concrete and F if all fixes were vague.
Coding was performed blind to cluster membership to prevent the coder from unconsciously steering responses toward fixes expected for a given archetype.

For statistical analysis, the 14 codes were collapsed into four categories:
general IT project management fixes (B1--B6 $\rightarrow$ B), segmentation-specific fixes (C1--C6 $\rightarrow$ C), emergent (E), and null (F).
Coding at the granular level (requiring the coder to identify the specific failure factor being addressed) improves accuracy at the collapsed level.
The most consequential coding error for the analysis would be confusing a B code with a C code; this requires a substantially larger misreading of the response than confusing, say, B1 with B3 or C4 with C5.
The collapsing step thus absorbs the most likely source of coding variability.

Table~\ref{tab:code_distribution} presents the distribution of codes across all 400 respondents.

Only one coder with the required domain expertise was available to the project, so independent coding by a second expert was not feasible.
To assess the stability of the Q8 coding under this constraint, the same coder coded all 400 responses a second time after a 120-day interval, blind to the original codes.
At the collapsed category level (B, C, E, F) used in the analysis, the two passes agreed on 90.2\% of responses, with Cohen's $\kappa = 0.789$, which corresponds to substantial agreement under the commonly cited benchmarks of Landis and Koch~\cite{landis1977measurement} and indicates that the coding is stable.

\section{Sample Free-form Text Responses} \label{sec:sample}

    \begin{quote}
        \itshape ``Make the project timeline more flexible so it can accommodate changes and iterations.''
    \end{quote}
    \begin{flushright}
        \footnotesize --- Perfect Storm Respondent
    \end{flushright}

    \begin{quote}
        \itshape ``Major coordination between network security and application teams can stop policy conflicts that broke key internal tools completely.''
    \end{quote}
    \begin{flushright}
        \footnotesize --- Diffuse Friction Respondent
    \end{flushright}

    \begin{quote}
        \itshape ``I would ensure the scope was strictly followed.''
    \end{quote}
    \begin{flushright}
        \footnotesize --- Operational Drag Respondent
    \end{flushright}

    \begin{quote}
        \itshape ``Better and smarter organizational leadership.''
    \end{quote}
    \begin{flushright}
        \footnotesize --- Scope \& Vis. Trap Respondent
    \end{flushright}

\newpage
\bibliographystyle{IEEEtran}
\bibliography{segfail}

\end{document}